\documentclass[pre,twocolumn,superscriptaddress,preprintnumbers,amsmath,amssymb]{revtex4-1}

\usepackage{epsfig}
\usepackage{amsmath}
\usepackage{amssymb}
\usepackage{amsfonts}
\usepackage{mathptmx}
\usepackage{dcolumn}
\usepackage{eucal}
\usepackage{bm}
\usepackage{color}
\usepackage[colorlinks,linkcolor=blue,citecolor=blue]{hyperref}
\usepackage{epstopdf}
\usepackage{bbold}
\usepackage{url}
\usepackage{enumerate}
\usepackage[table,xcdraw]{xcolor}

\usepackage{dsfont}

\newcommand{\bs}{\boldsymbol}

\newcommand{\eps}{\epsilon}

\newcommand{\hatstate}{\hat{\rho}}
\newcommand{\hatSig}{\hat{\sigma}}
\newcommand{\average}[1]{\langle #1 \rangle}
\newcommand{\Trace}{{\rm Tr}}

\newcommand{\hatH}{\hat{H}}

\newcommand{\hatP}{\hat{P}}

\newcommand{\Prob}{\mathcal{P}}

\def\be{\begin{eqnarray}}
\def\ee{\end{eqnarray}}

\begin{document}

\title{Leggett-Garg inequalities for quantum fluctuating work}

\author{H. J. D. Miller}
\email{hm419@exeter.ac.uk}
\author{J. Anders}
\email[]{janet@qipc.org}
\affiliation{Department of Physics and Astronomy, University of Exeter, Stocker Road, EX4 4QL, United Kingdom.}
\date{\today}

\begin{abstract}
The Leggett-Garg inequalities serve to test whether or not quantum correlations in time can be explained within a classical macrorealistic framework. We apply this test to thermodynamics and derive a set of Leggett-Garg inequalities for the statistics of fluctuating work done on a quantum system unitarily driven in time. It is shown that these inequalities can be violated in a driven two-level system, thereby demonstrating that there exists no general macrorealistic description of quantum work. These violations are shown to emerge within the standard Two-Projective-Measurement scheme as well as for alternative definitions of fluctuating work that are based on weak measurement. Our results elucidate the influences of temporal correlations on work extraction in the quantum regime and highlight a key difference between quantum and classical thermodynamics. 
\end{abstract}

\date{\today}

\maketitle

\section{Introduction}

Much like the celebrated Bell inequalities, which shed light on the deeply non-classical properties of spatial correlations encountered in entangled systems, quantum mechanics posseses a rich temporal structure that distinguishes it from classical physics. In 1985 Leggett and Garg explored this structure by introducing the concept of \textit{macrorealism} \cite{Leggett1985}. In essence macrorealism can be condensed into two main assumptions about the temporal properties of physical observables within any \textit{classical} description of physics \cite{Leggett1985,Kofler2008,Williams2008,Wilde2012,Emary2014,Kofler2013a,Halliwell2015a,Clemente2015}; 
\begin{enumerate}[(i)]
\item \textit{Macrorealism per se}: physical observables take on well-defined values at all times independent of the act of observation.
\item \textit{Non-invasive measurability}: in principle it is possible to measure the value of an observable without changing the subsequent evolution of the system.
\end{enumerate}
Assumptions (i) and (ii) can be used to derive mathematical inequalities, the so-called Leggett-Garg inequalities, that serve to test the macrorealism of physical observables. Violations of these inequalities subsequently rule out what would be expected in a classical system, and this quantum behaviour has now been confirmed experimentally in a variety of settings \cite{palacio2010,dressel2011,Goggin2011a,Knee2012,Zhou2012,Robens2015a}.

In the quantum regime, thermodynamic quantities such as fluctuating work and heat cannot be represented by hermitian observables, but are conventionally defined via multi-time projective measurements performed on the system \cite{Talkner2007c,Esposito2009,Hanggi2015}. For a closed quantum system, one way of defining the fluctuating work done on the system driven out of equilibrium is by the difference in energy eigenvalues observed at the start and end of its evolution. This framework is  commonly referred to as the two-projective-measurement scheme, and serves as a route to many of the known fluctuation theorems such as the Jarzynski equality \cite{Talkner2007c,Esposito2009} and Tasaki-Crooks relation \cite{Talkner2007a}. Given that these results mirror the corresponding classical fluctuation relations \cite{Jarzynski1997d,Crooks}, it is often assumed that work is simply a classical stochastic variable even within the quantum regime. However, the influence of non-classical temporal correlations that arise from two-time quantum measurements on the statistics of fluctuating work have yet to be fully understood. One aspect of this is the fact that work measurements remove coherences in the energy basis and can affect the future evolution of the quantum system, modifying the average work done during the process \cite{Kammerlander2016,Miller2017}. Alternative definitions of quantum work related to weak measurement have been proposed in order to circumvent this effect of measurement disturbance on the statistics of work \cite{Allahverdyan2014c,Solinas2015,Miller2017}. However, it has been shown that the resulting quantum work distributions are not generally positive-definite \cite{Baumer2017}. The emergence of negative quasi-probabilities is a signature of quantum behaviour, and hints at a link to violations of the Leggett-Garg inequalities \cite{Bednorz2012b,Halliwell2015a,Hofer2017a}. In a similar vein, violations of macrorealism have also been related to the presence of anomalous weak values in quantum systems \cite{Williams2008}. Recent work by Blattmann and M\o lmer \cite{Blattmann2017} has successfully linked violations of macrorealism to quantum work in the standard TPM approach by utilising the entropic Leggett-Garg inequalities. In their approach one compares the Shannon entropy of the work distribution over different intervals of time. However, the Shannon entropy is not well-defined if the work distribution fails to be positive, and so the entropic Leggett-Garg inequalities cannot be applied to situations in which the work distribution becomes a quasi-probability. 

In this paper we will utilise the assumptions of Leggett and Garg to demonstrate that there exists no general macrorealistic description of work for quantum systems driven out of equilibrium. In particular, we show that quantum temporal correlations between energy measurements performed at different times influence the statistical moments of the fluctuating work done on the system during a non-equilibrium process. This result is shown to hold for three different definitions of quantum work: the two-projective measurement (TPM) scheme \cite{Talkner2007c}, the full-counting statistics (FCS) \cite{Solinas2015} and the Margenau-Hill (MH) work distribution \cite{Allahverdyan2014c}. Crucially the inequalities that we derive can be used to test for violations of macrorealism in both strong and weak measurement schemes, regardless of whether or not the work distribution is positive or not.

The paper is organised as follows: we first introduce a set of Leggett-Garg inequalities for the moments of fluctuating work, and then consider a driven two-level system and show that the inequalities can be violated. Following that we introduce an alternative set of Leggett-Garg inequalities for the moment-generating function, and apply these inequalities to alternative definitions of quantum work that are based on weak measurement, namely the FCS and MH definitions, subsequently showing that violations of macrorealism can also occur. Finally we conclude with a discussion of our results.

\section{Inequalities for moments of work}\label{sec:momentsineq}

We first recall the setup for the original Leggett-Garg inequalities \cite{Leggett1985}. First consider performing three protocols in which the spin $S(t_{i})=S_{i}=\pm1$ of a qubit is projectively measured at two times within a set of three times $t_0<t_1<t_2$. For each of the three protocols one can obtain the temporal correlation function for the values of the spin at times $t=t_i$ and $t=t_j$, denoted $C_{ij}=\average{S_iS_j}$. The macrorealism assumptions (i) and (ii) imply that there exists a three-time probability distribution $\Prob(S_0,S_1,S_2)$ such that the distributions describing the statistics of each individual protocol can be obtained as marginals of this three-time distribution, eg. $\Prob(S_0,S_2)=\sum_{S_1}\Prob(S_0,S_1,S_2)$ and so on \cite{Emary2014}. Note that while assumption (i) implies the existence of a three-time probability with the correct marginals, assumption (ii) guarantees that this distribution is the same for all three separate experiments \cite{Emary2014}. Finally, using the marginal properties of $\Prob(S_0,S_1,S_2)$ yields the following Leggett-Garg inequality relating the correlation functions for the three protocols \cite{Leggett1985};

\be\label{leg}
	C_{01}+C_{12}-C_{02} \leq 1.
\ee
	
This inequality holds for any dichotomic observable. For a simple qubit, the spin at time $t_{i}$ can be represented by a combination of Pauli matrices; $\hat{S}_{i}=\boldsymbol{s}_{i}\cdot\boldsymbol{\hat{\sigma}}$. By performing successive projective measurements of the spin, the correlation functions can be obtained for each of the three protocols. This leads to a violation the RHS of Eq.~(\ref{leg}) which can take a maximum value of $\frac{3}{2}$ \cite{Emary2014}. This example illustrates the failure of macrorealism for quantum systems. 

\begin{figure}[t]
\includegraphics[scale=0.4]{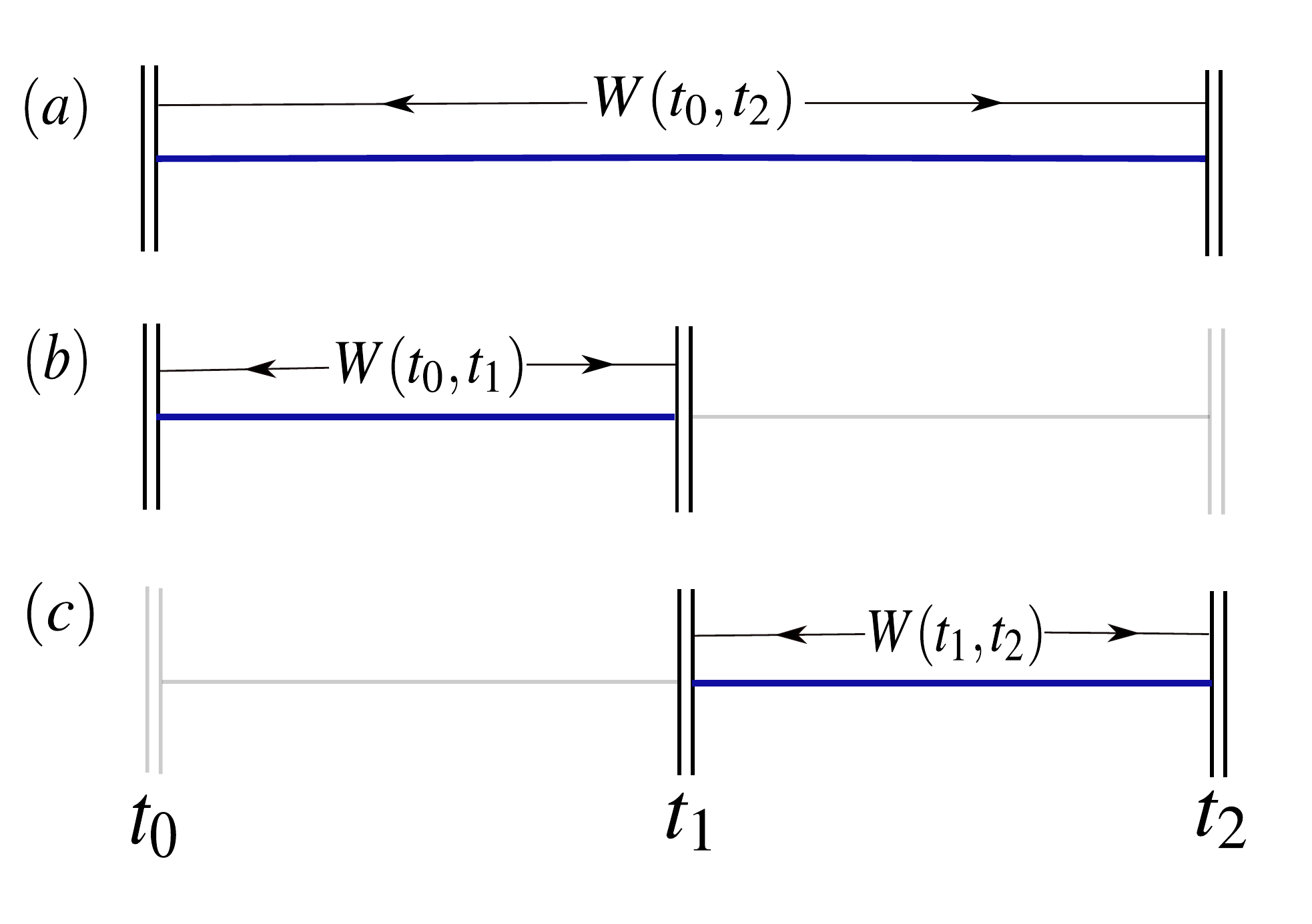}
\caption{\label{fig:setup} \textbf{Schematic diagram for detecting non-classical work statistics.} An observer performs three separate experiments, (a)-(c), in which the fluctuating work done on the system is measured between the time intervals shown in the diagram. To test the validity of the Leggett-Garg inequality for work, Eq.~(\ref{legineq}), one compares the statistics of the three experiments, with the same initial state chosen at time $t_0$. Note that in experiment (c) no measurement is made at $t_0$, thus the system evolves unitarily up to time $t_1$.}
\end{figure}

It is also possible to derive a set of Leggett-Garg inequalities reminiscent of Eq.~(\ref{leg}) for the moments of fluctuating work in a closed quantum system driven in time. For simplicity we consider a system that can occupy one of two fixed energy states, which we denote by $\frac{\eps}{2}$ and $-\frac{\eps}{2}$, at three points in time $t=t_0<t_1<t_2$ during the driving process. For the statistics of work measured during a particular time interval $t\in [ t_i,t_j ]$, the $k$'th moment of fluctuating work is defined as 
\be\label{moments}
	\average{W^{k}(t_i,t_j)}=\sum_{\eps_i,\eps_j}\Prob(\eps_i,\eps_j)(\eps_j-\eps_i)^{k}.
\ee
Here the energies occupied by the system at time $t_i$ are denoted by $\eps_i$ and we assume that the possible work values are given by the energy changes $W_{ij}=\eps_j-\eps_i$. The probability $\Prob(\eps_i,\eps_j)$ governs the statistics of energy at times $t_i$ and $t_j$, and we make no assumptions about the exact definition of $\Prob(\eps_i,\eps_j)$ aside from assuming it is normalised and non-negative. As with the standard Leggett-Garg experiment described above, the aim is to compare the work statistics observed within different time intervals along the driving process, as shown in Figure~\ref{fig:setup}. Thus in analogy with Eq.~(\ref{leg}) we will consider the following quantity;
\be\label{legwork}
	M_k=\average{W^{k}(t_0,t_{1})}+\average{W^{k}(t_1,t_2)}-\average{W^{k}(t_0,t_2)}.
\ee
This quantity can be measured over many runs of the driving process during each of the three time intervals, where the system is prepared in same state at time $t_0$ for each experiment. For example $M_1$ is obtained by measuring the average work done on the system sequentially during intervals $t\in [ t_0,t_1 ]$ and $t\in [ t_1,t_2 ]$, and then subtracting the average work done during the total time interval $t\in [ t_0,t_2 ]$. We now seek to bound Eq.~(\ref{legwork}) through the assumption that the fluctuating work is a macrorealistic variable, as defined by (i) and (ii).

These assumptions imply the existence of a global probability distribution $\Prob(\eps_0,\eps_1,\eps_2)$ describing the energy statistics at all points in time along the driving process, where each two-time distribution can be obtained as a marginal;
\be\label{marginal}
	\Prob(\eps_i,\eps_j)=\sum_{m\neq i,j}\Prob(\eps_0,\eps_1,\eps_{2}) \ \ \ \forall i,j.
\ee
We can show that this condition immediately leads to the following Leggett-Garg inequality for the moments of work (see Appendix~\ref{sec:a0});
\be\label{legineq}
	\nonumber&M_k\geq 0, \ \ \ \text{even} \ k, \\
	 &M_k = 0, \ \ \ \text{odd} \ k.
\ee
For $k=1$ this implies $\average{W(t_0,t_{1})}+\average{W(t_1,t_2)}=\average{W(t_0,t_2)}$. This makes intuitive sense; in classical thermodynamics one would not expect to observe any difference between the sum of each intermediate average amount work done and the total average work done between the initial and final points in time. However, we will subsequently show that this does not generally hold for quantum systems, as the bounds in Eq.~(\ref{legineq}) can be violated for certain driving processes. It should also be noted that while we have assumed a discrete energy spectrum for the time-dependent Hamiltonian, this is not crucial to the derivation of Eq.~(\ref{legineq}). Indeed, so long as one assumes that the energy moments are always finite then inequalities of the form Eq.~(\ref{legineq}) can be derived. However, we will restrict our attention to a two-dimensional quantum system throughout the paper for simplicity, as this is sufficient to demonstrate violations of macrorealism in the statistics of fluctuating work.

\pagebreak

\section{Violations of the Leggett-Garg inequalities for work moments}\label{sec:legmoments}

We will now utilise the inequalities Eq.~(\ref{legineq}) to show that fluctuating work can not generally be described by a macrorealist theory for quantum systems. The relevant situation that we consider is a standard setup for the thermodynamics of work extraction; an isolated system is initially thermalised and then driven out of equilibrium via changing its Hamiltonian in time, extracting work in the process \cite{Talkner2007c,Esposito2009,Hanggi2015}. Suppose that we have a two-level system described by a time-dependent Hamiltonian $\hatH_H(t)$ in the Heisenberg picture with initial state $\hatstate$ such that;
\be\label{ham}
	\hatH_H(t_i)=\frac{\eps}{2}\bs{a}_{i}\cdot \bs{\hatSig};  \ \ \ \ \ \ \ \ \hatstate=\frac{1}{2}(\hat{I}+\bs{r}\cdot\bs{\hatSig}),
\ee
where $|\bs{r}|\leq1$ and $|\bs{a}_{i}|=1$ are vectors. Note that the time-dependence of the Hamiltonian is attributed only to each vector $\bs{a}_{i}$. In an experimental setup Eq.~(\ref{ham}) describes a spin-1/2 particle coupled to an external classical magnetic field, with the direction of the applied field adjusted by the experimenter in time. Without loss of generality we will set the initial Hamiltonian along the z-axis of the Bloch sphere, i.e. $\bs{a}_{0}=\lbrace 0,0,1 \rbrace$, and choose an initial thermal state with respect to $\hatH(t_0)$ at inverse temperature $\beta$; $\hatstate \propto \text{exp}(-\beta \hatH(t_0))$. This in turn implies that $\bs{r}=\lbrace0,0,-\text{tanh}(\beta \eps /2)\rbrace$. To obtain the moments of work in Eq.~(\ref{legwork}) two projective energy measurements are performed at the start and end of the driving process within the fixed time intervals shown in Figure~\ref{fig:setup}. This method is commonly referred to as the two-projective measurement scheme (TPM) \cite{Talkner2007c,Esposito2009,Jarzynski2015a}. The joint probability to observe energy $\eps_{i}$ at $t=t_i$ and then $\eps_{j}$ at $t=t_j$ is given by
\be\label{wigner}
	\Prob\bigg(\eps_i=\pm\frac{\eps}{2},\eps_j=\pm\frac{\eps}{2}\bigg)=\Trace\big[\hatstate\hatP^{\pm}_{a_i}\big]\cdot\Trace\big[\hatP^{\pm}_{a_j}\hatP^{\pm}_{a_i}\big].
\ee
Here we have denoted $\hatP^{\pm}_{a_i}$ as the projector onto the relevant energy state of the Hamiltonian at time $t_i$. Substituting Eq.~(\ref{wigner}) into Eq.~(\ref{moments}) gives the moments of work from successive projective energy measurements (see Appendix~\ref{sec:a}):
\be\label{momentsquantum}
	\average{W^{k}(t_i,t_j)}=\begin{cases}
        (1-\bs{a}_j\cdot\bs{a}_i)\frac{\eps^{k}}{2}, & \text{even k,}\\[1em]
        -(\bs{r}\cdot\bs{a}_{i})(1-\bs{a}_j\cdot\bs{a}_{i})\frac{\eps^{k}}{2}, & \text{odd k}.
  \end{cases}
\ee
To identify the conditions under which fluctuations in work violate Eq.~(\ref{legineq}), we substitute the above expression into Eq.~(\ref{legwork}) and parameterise the driving process by introducing $\text{cos}(\theta_{ij})=\bs{a}_j\cdot\bs{a}_{i}$. For even $k$ we have the following quantum bound for $M_k$ after minimising over all normalised vectors $\lbrace \bs{a}_{i} \rbrace$ for $i=0,1,2$ (see Appendix~\ref{sec:a}); 
\be\label{legeven2}
	\min_{\lbrace \bs{a}_{i} \rbrace}\bigg[M_k \bigg]= -\frac{\eps^{k}}{4}; \ \ \ \ \textit{(even k)}.
\ee
which is saturated by choosing $\theta_{10}=\theta_{21}=\pi/3$. Secondly, for odd $k$ we find that maximising over $\lbrace \bs{a}_{i} \rbrace$ gives
\be\label{legodd2}
	\max_{\lbrace \bs{a}_{i} \rbrace}\bigg[\big| M_k\big| \bigg]= \bigg|\frac{\eps^{k}}{2}\text{tanh}(\beta \eps/2)\bigg|; \ \ \ \ \textit{(odd k)},
\ee
where the bound is saturated by choosing $\theta_{10}=\theta_{21}=\pi/2$. From Eq.~(\ref{legeven2}) and Eq.~(\ref{legodd2}) it is now apparent that there exists unitary protocols that violate the bounds in Eq.~(\ref{legineq}). Given that the bounds Eq.~(\ref{legineq}) necessarily follow from assumptions (i) and (ii), we conclude that quantum fluctuating work generally lacks a macrorealistic description.

Let us note that Eq.~(\ref{legineq}) cannot be violated for odd $k$ in the high temperature limit. However, Eq.~(\ref{momentsquantum}) shows that the even work moments are independent of $\beta$, and so Eq.~(\ref{legineq}) can indeed be violated for even $k$ regardless of temperature. This is not surprising, as the standard Leggett-Garg inequality can be violated in the case of an initially maximally mixed qubit \cite{Emary2014}. Violations of the work Leggett-Garg inequality, Eq.~(\ref{legineq}), can still occur at all temperatures since the system may acquire coherences in energy at intermediate times due to the unitary driving.

\section{Inequalities for the characteristic function of work}\label{sec:mgf}

While the inequalities Eq.~(\ref{legineq}) provide a simple identification of non-classicality for the moments of work, we will show in this section that it is possible to condense this information into two inequalities related to the characteristic function for work rather than the moments themselves. A similar method has previously been applied to investigate the non-classical properties of electron-transport through conductors \cite{Emary2012a}. The benefit of this approach is two-fold. While the projective energy measurements used to obtain the moments in Eq.~(\ref{moments}) may be difficult to implement in practice, measurements of the characteristic function for work can be performed via ancilla-assisted measurement as shown in \cite{Dorner2013b,Mazzola2013a,Roncaglia2014c,Batalhao2014}. Secondly, inequalities for the characteristic function allow us to consider alternative non-invasive measurement schemes such as the full-counting statistics approach to the quantum work distribution proposed in \cite{Solinas2015}, as we later show in the next section.

The characteristic function uniquely defines the probability distribution for work in a unitarily driven system, and is obtained through the Fourier transform of the work distribution;
\be\label{mgf}
	G_{\lambda}(t_i,t_j)=\sum_{\eps_i,\eps_j}\Prob(\eps_i,\eps_j)e^{i\lambda(\eps_j-\eps_i)},
\ee 
with work values $\eps_j-\eps_i$. Assuming the same protocol given by Eq.~(\ref{ham}), we can consider a linear combination of characteristic functions for the three intervals of time shown in Figure~\ref{fig:setup};
\be\label{legmgf}
	L_{\lambda}=G_{\lambda}(t_0,t_1)+G_{\lambda}(t_1,t_2)-G_{\lambda}(t_0,t_2).
\ee
The assumptions (i) and (ii) for macrorealism imply the following upper bound on the real part of Eq.~(\ref{legmgf}) for the qubit system (see Appendix~\ref{sec:b});
\be\label{realbound}
	\mathcal{R}\text{e}(L_{\lambda}) \leq 1,
\ee
whilst the imaginary part of Eq.~(\ref{legmgf}) becomes an \textit{equality};
\be\label{imbound}
	\mathcal{I}\text{m}(L_{\lambda})=0. 
\ee
The bounds Eq.~(\ref{realbound}) and Eq.~(\ref{imbound}) now constitute a pair of Leggett-Garg-type inequalities for any classical characteristic function for fluctuating work, assuming that the energy is given by either $\frac{\eps}{2}$ or $-\frac{\eps}{2}$ at all times along the driving protocol.

Within the TPM scheme the characteristic function for work is given by \cite{Talkner2007c,Esposito2009};
\be\label{mgf}
	G^{\text{TPM}}_{\lambda}(t_i,t_j)=\Trace\big[\hat{\eta}_{i} \ e^{i\lambda\hatH_H(t_j)}e^{-i\lambda\hatH_H(t_i)}\big],
\ee
where $\hat{\eta}_{i}$ is the initial state $\rho$ decohered in the basis of $\hatH(t_i)$. The quantum upper bound for the real part of Eq.~(\ref{legmgf}) is as follows (see Appendix~\ref{sec:c});
\be\label{qmfgreal}
	\max_{\lbrace \bs{a}_{i} \rbrace}\bigg[\mathcal{R}\text{e}(L_{\lambda})\bigg] =\frac{5}{4}-\frac{1}{4}\text{cos}(\lambda \eps),
\ee
Consequently the upper bound exceeds the classical inequality Eq.~(\ref{realbound}) for all $\lambda$. Secondly, the upper bound for the imaginary part of Eq.~(\ref{legmgf}) is
\be\label{qmgfim}
	\max_{\lbrace \bs{a}_{i} \rbrace}\bigg[\big|\mathcal{I}\text{m}(L_{\lambda})\big|\bigg]= \bigg| \frac{1}{2}\text{tanh}(\beta \eps/2)\text{sin}(\lambda \eps)\bigg|,
\ee
with a maximum violation of $\mathcal{I}\text{m}(L^{G}_{\lambda})=\frac{1}{2}\text{tanh}(\beta \eps/2)$ for $\lambda=\pi/2\eps$. As one would expect, the quantum bounds Eq.~(\ref{qmfgreal}) and Eq.~(\ref{qmgfim}) are obtained by choosing the same protocol-dependent parameters used to obtain Eq.~(\ref{legeven2}) and Eq.~(\ref{legodd2}) respectively. 

To summarise this section we have presented a Leggett-Garg inequality for the characteristic function of fluctuating work, and shown that the non-classicality observed in the moments of work is also exhibited in the characteristic function itself for the TPM scheme.    

\section{Generalisation to weak measurements of work}\label{sec:fcs}

In the previous section we simply recast the original violations of the Leggett-Garg inequality Eq.~(\ref{legineq}) into the form relevant to the characteristic function for the work statistics of a qubit. While this was applied to the TPM protocol, the inequalities Eq.~(\ref{realbound}) and Eq.~(\ref{imbound}) apply to any qubit with a fixed energy spectrum, and this ultimately allows us to investigate alternative measurement schemes and their resulting non-classical violations. Due to the invasive nature of projective measurements, some have argued that the definition of fluctuating work is thermodynamically inconsistent when applied to states with initial coherences in energy \cite{Allahverdyan2014c,Kammerlander2016,Solinas2015,Baumer2017}. In turn this has inspired formulations of non-invasive work statistics that remain consistent with energy conservation in closed systems \cite{Allahverdyan2014c,Solinas2015,Miller2017}. In particular, we will consider the full-counting statistics for fluctuating work \cite{Nazarov2003,Clerk2011,Bednorz2012b,Emary2012a,Hofer2015,Solinas2015}. To obtain the full-counting statistics for work, one couples the system's Hamiltonian to the momentum of an external detector, and subsequently measures the phase change acquired by the detector's momentum during the driving process given by the system's time-dependent Hamiltonian $H(t)$ \cite{Clerk2011}. In turn this allows one to reconstruct a characteristic function of the following form \cite{Solinas2015,Solinas2016a};
\be\label{eq:fcs}
	G^{\text{FCS}}_{\lambda}(t_i,t_j)=\Trace\big[\hatstate e^{-i\frac{\lambda}{2}\hatH_H(t_i)}e^{i\lambda\hatH_H(t_j)}e^{-i\frac{\lambda}{2}\hatH_H(t_i)}\big].
\ee
For example, as shown in \cite{Solinas2015} the first moment of Eq.~(\ref{eq:fcs}) yields $\average{W(t_i,t_j)}=\average{H(t_{j})}-\average{H(t_{i})}$; in other words the average work done on the system is simply the difference in average energy evaluated in the Heisenberg picture. Consequently one can easily see that the first moment in this measurement scheme still obeys the Leggett-Garg inequality Eq.~(\ref{legineq}). However, we demonstrate that the same is not true for the higher order moments contained in the characteristic function Eq.~(\ref{eq:fcs}). Considering again the quantum two-level system described by Eq.~(\ref{ham}), it can be shown that the real part of Eq.~(\ref{eq:fcs}) is equal to the real part of Eq.~(\ref{mgf}) obtained via the TPM scheme (see Appendix~\ref{sec:d});
\be\label{eq:fcsreal}
	\mathcal{R}\text{e}\bigg(G^{\text{FCS}}_{\lambda}(t_i,t_j)\bigg)=\mathcal{R}\text{e}\bigg(\Trace\big[\hat{\eta}_{i} \ e^{i\lambda\hatH_H(t_j)}e^{-i\lambda\hatH_H(t_i)}\big]\bigg),
\ee
This is a surprising result, as it suggests that the same violations of Eq.~(\ref{realbound}), ie. the upper bound Eq.~(\ref{qmfgreal}), can be obtained non-invasively. It is only the imaginary part of $G^{\text{FCS}}_{\lambda}(t_i,t_j)$ that differs from the TPM scheme, in which we find the following upper bound for Eq.~(\ref{legmgf});
\be\label{eq:fcsim}
	\max_{\lbrace \bs{a}_{i}\rbrace}\bigg[\big|\mathcal{I}\text{m}(L^{FCS}_{\lambda})\big|\bigg]=\bigg|2\text{sin}^{2}\big(\frac{\lambda \eps}{4}\big)\text{sin}\big(\frac{\lambda \eps}{2}\big)\text{tanh}\big(\frac{\beta \eps}{2}\big)\bigg|
\ee 
which is again obtained by choosing $\theta_{10}=\theta_{21}=\pi/2$. 


The full-counting statistics are not the only way to characterise non-invasive measurements of work. As proposed by Allahverdyan \cite{Allahverdyan2014c}, an alternative characteristic function describing the statistics of work derived from the Margenau-Hill distribution for successive energy measurements \cite{Margenau1961,Johansen2007} is as follows (see Appendix~\ref{sec:e}): 
\be\label{eq:mh}
	G^{\text{MH}}_{\lambda}(t_i,t_j)=\Trace\big[\hatstate e^{i\lambda\hatH_H(t_j)}\star e^{-i\lambda\hatH_H(t_i)}\big],
\ee
\newline
where $\hat{A}\star\hat{B}=\frac{1}{2}[\hat{A}\hat{B}+\hat{B}\hat{A}]$ denotes the symmetric Jordan product. Notably the corresponding probability distribution can be obtained via sequential weak measurement \cite{Lundeen2012}. While the first and second moments, $\average{W}$ and $\average{W^2}$, are the same as those obtained from the full-counting statistics, in general higher order moments differ. However, for the isolated driven qubit we again find precisely the same violations of Eq.~(\ref{realbound}) because the real part of $G^{\text{MH}}_{\lambda}(t_i,t_j)$ is also equivalent to the real part of $G^{\text{FCS}}_{\lambda}(t_i,t_j)$. On the other hand the inequality for the imaginary term, Eq.~(\ref{imbound}), cannot be violated in our setup (Appendix~\ref{sec:e}).
\bigskip

\section{Discussion}\label{sec:disc}

In the paper we have demonstrated a violation of macrorealism in the statistics of fluctuating work for a quantum system unitarily driven out of thermal equilibrium for three different characterisations of the work statistics. As with the original Leggett-Garg inequalities, these violations stem from the absence of a global three-time probability distribution of the form Eq.~(\ref{marginal}) for both strong and weak measurement schemes. This emphasises the fact that quantum fluctuations in work are manifestly different from the stochastic fluctuations encountered in classical non-equilibrium thermodynamics due to the influence of temporal correlations on the work moments. These findings compliment recent results showing that fluctuating work cannot always be assigned a well-defined probability distribution when quantum coherence is taken into account \cite{Allahverdyan2014c,Solinas2015,Miller2017,Solinas2017,Baumer2017,Lostaglio2017}. Ultimately our analysis shows that the Leggett-Garg inequalities provide a useful tool for understanding the difference between quantum and classical thermodynamics, and the inequalities Eq.~(\ref{legineq}), Eq.~(\ref{realbound}) and Eq.~(\ref{imbound}) may find an application in identifying quantum behaviour in thermal machines \cite{Friedenberger2015}.

\


\begin{acknowledgments}
We would like to thank Paolo Solinas for suggesting a link between quantum work and the Leggett-Garg inequality, and Clive Emary and Alexander Friedenberger for insightful discussions. HM is supported by EPSRC through a Doctoral Training Grant. J.A. acknowledges support from EPSRC, grant EP/M009165/1, and the Royal Society. This research was  supported by the COST network MP1209 ``Thermodynamics in the quantum regime".
\end{acknowledgments}

\bibliographystyle{apsrev4-1}
\bibliography{LGbib.bib}

\appendix

\widetext

\section{Derivation of Eq.~(\ref{legineq})}\label{sec:a0}

\noindent By using the condition Eq.~(\ref{marginal}) for the marginal probabilities describing the statistics of the three experiments displayed in Figure~(\ref{fig:setup}), one can express the following linear combination of work, Eq.~(\ref{legwork}) as;
\be\label{legwork2}
	\nonumber M_k&=&\sum_{\eps_0, \eps_1, \eps_2}\Prob(\eps_0, \eps_1, \eps_2)\bigg[(\eps_{2}-\eps_{1})^{k}+(\eps_{1}-\eps_{0})^{k}
	-(\eps_{2}-\eps_{0})^{k} \bigg] \\
	&=&\bigg(\eps^{k}+(-1)^{k}\eps^{k}\bigg).\bigg[\Prob(+, -, +)+\Prob(-, +, -) \bigg]
\ee
Note here that $\Prob(+,-,+)$ refers to the sequence $\big\lbrace\eps_0=+\eps/2,\eps_1=-\eps/2,\eps_{2}=+\eps/2\big\rbrace$ and similarly for $\Prob(-,+,-)$. The last line in Eq.~(\ref{legwork2}) is reached by using the fact $\eps_i=\pm\eps/2, \ \ \  \forall i$ and that all other sequences drop out of the above summation regardless of the probability. Given that one is free to choose any arbitary distribution $\Prob(\eps_0,\eps_1,\eps_2)$, we arrive at a Leggett-Garg inequality for fluctuating work;
\be
	0\leq M_k \leq \eps^{k}+(-1)^{k}\eps^{k}, \ \ \ \forall k.
\ee
where the lower bound is achieved by choosing $\Prob(+, -, +)=\Prob(-, +, -)=0$ and the upper bound by setting $\Prob(+, -, +)=\Prob(-, +, -)=\frac{1}{2}$.

\section{Derivation of Eq.~(\ref{momentsquantum})}\label{sec:a}

\noindent We begin with the standard formula for the joint probability to observe energy $\eps_i=\pm\frac{\eps}{2}$ at time $t_i$ and then energy $\eps_j=\pm\frac{\eps}{2}$ at time $t_j$ from projective measurements of the Hamiltonian;
\be
	\Prob\bigg(\eps_i=\pm\frac{\eps}{2},\eps_j=\pm\frac{\eps}{2}\bigg)=\Trace\big[\hatstate\hatP^{\pm}_{a_i}\big].\Trace\big[\hatP^{\pm}_{a_j}\hatP^{\pm}_{a_i}\big].
\ee
The above formula can be simplified by using the fact that each projector can be written as $\hatP^{\pm}_{a_i}=\frac{1}{2}(\hat{I}\pm \bs{a}_{i}\cdot \bs{\hatSig} )$, and then applying the identity $\Trace[(\bs{a}\cdot \bs{\hatSig})(\bs{b}\cdot \bs{\hatSig})]=2\bs{a}\cdot\bs{b}$. This leads to
\be
	\nonumber\Prob\bigg(\eps_i=-\frac{\eps}{2},\eps_j=+\frac{\eps}{2}\bigg)&=&\frac{1}{16}\Trace\bigg[(\hat{I}+\bs{r}\cdot\bs{\hatSig}).(\hat{I}-\bs{a}_{i}\cdot\bs{\hatSig})\bigg].\Trace\bigg[(\hat{I}+\bs{a}_{j}\cdot\bs{\hatSig}).(\hat{I}-\bs{a}_{i}\cdot\bs{\hatSig})\bigg]\\
	&=&\frac{1}{4}\bigg[(1-\bs{r}\cdot\bs{a}_i).(1-\bs{a}_j\cdot\bs{a}_i)\bigg],	
\ee
and similarly
\be
\Prob\bigg(\eps_i=+\frac{\eps}{2},\eps_j=-\frac{\eps}{2}\bigg)=\frac{1}{4}\bigg[(1+\bs{r}\cdot\bs{a}_i).(1-\bs{a}_j\cdot\bs{a}_i)\bigg]
\ee
Finally from the definition of the moments for work, Eq.~(\ref{moments}), we arrive at Eq.~(\ref{momentsquantum});
\be
	\nonumber \average{W^{k}(t_i,t_j)}&=&\sum_{\eps_i,\eps_j}\Prob(\eps_i,\eps_j)(\eps_j-\eps_i)^{k} \\
	\nonumber&=&\eps^{k}\big[\Prob\big(-,+\big)+(-1)^{k}\Prob\big(+,-\big) \big] \\
	&=&\frac{(1-\bs{a}_j\cdot\bs{a}_i)}{4}\eps^{k}\bigg[1-\bs{r}\cdot\bs{a}_{i}+(-1)^{k}(1+\bs{r}\cdot\bs{a}_{i})\bigg].
\ee
Now the moments can be substituted into the Leggett-Garg equation Eq.~(\ref{legwork}). For even $k$ the inequality is state independent and can be rewritten in terms of the angles between each Bloch vector, denoted $\bs{a}_i\cdot\bs{a}_{j}=\text{cos}(\theta_{ij})$;
\be\label{legeven}
	M_k=\frac{\eps^{k}}{2}\bigg[1+\text{cos}(\theta_{10}+\theta_{21})-\text{cos}(\theta_{21})-\text{cos}(\theta_{10})\bigg]
\ee
where we have used the trigonometric relation $\theta_{10}+\theta_{21}=\theta_{20}$. The lower bound, Eq.~(\ref{legeven2}) is obtained by setting the angles to $\theta_{10}=\theta_{21}=\pi/3$. To bound Eq.~(\ref{legwork}) for odd $k$ we note the following relation; $\bs{r}\cdot\bs{a}_{1}=\text{tanh}(\beta \eps/2)\text{cos}(\theta_{01})$. In this case Eq.~(\ref{legwork}) can be be written as follows;
\be\label{legodd}
	M_k=\frac{\eps^{k}}{2}\text{tanh}(\beta \eps/2) \bigg[\text{cos}(\theta_{01})\text{cos}(\theta_{12})-\text{cos}(\theta_{01}+\theta_{12})\bigg]
\ee
The bound Eq.~(\ref{legodd2}) is obtained by setting $\theta_{10}=\theta_{21}=\pi/2$.

\section{Derivation of Eq.~(\ref{realbound}) and Eq.~(\ref{imbound})}\label{sec:b}

The characteristic function for work done by a system driven unitarily in time between $t=t_i$ and $t=t_j$ is given by the following;
\be
	G(\lambda,t_i,t_j)&=&\average{e^{i\lambda W(t_i,t_j)}}, \\
	&=&\sum_{\eps_i,\eps_j}\Prob(\eps_i,\eps_j)e^{i\lambda(\eps_j-\eps_i)},
\ee
where $\Prob(\eps_i,\eps_j)$ is some arbitrary joint probability governing the statistics of energy at two separate times. Under the assumptions (i) and (ii) for macrorealism this again implies the existence of a three-time distribution of the form Eq~.(\ref{marginal}), meaning that Eq.~(\ref{legmgf}) can be rewritten as
\be
	L_{\lambda}&=&G(\lambda,t_1,t_2)+G(\lambda,t_0,t_1)-G(\lambda,t_0,t_2) \\
	&=&\sum_{\eps_1,\eps_2}\Prob(\eps_1,\eps_2)e^{i\lambda(\eps_2-\eps_1)}+\sum_{\eps_0,\eps_1}\Prob(\eps_0,\eps_1)e^{i\lambda(\eps_1-\eps_0)}-\sum_{\eps_0,\eps_2}\Prob(\eps_0,\eps_2)e^{i\lambda(\eps_2-\eps_1)} \\
	&=&\sum_{\eps_0, \eps_1, \eps_2}\Prob(\eps_0, \eps_1, \eps_2)\bigg[e^{i\lambda(\eps_{2}-\eps_1)}+e^{i\lambda(\eps_{1}-\eps_0)}-e^{i\lambda(\eps_{2}-\eps_0)} \bigg]	.
\ee
Defining $\theta_{ij}=\lambda(\eps_{j}-\eps_i)$, the real and imaginary parts of Eq.~(\ref{legmgf}) are then
\be\label{reim}
	\nonumber&\mathcal{R}\text{e}&(L_{\lambda})=\sum_{\eps_0, \eps_1, \eps_2}\Prob(\eps_0, \eps_1, \eps_2)\bigg[\text{cos}(\theta_{12})+\text{cos}(\theta_{01})-\text{cos}(\theta_{01}+\theta_{12})\bigg] \\
	&\mathcal{I}\text{m}&(L_{\lambda})=\sum_{\eps_0, \eps_1, \eps_2}\Prob(\eps_0, \eps_1, \eps_2)\bigg[\text{sin}(\theta_{12})+\text{sin}(\theta_{01})-\text{sin}(\theta_{01}+\theta_{12})\bigg]
\ee
To obtain bounds on the real and imaginary terms in Eq.~(\ref{reim}) we use the fact that $\eps_{i}=\pm \frac{\eps}{2}$ at all times and consider the various combinations of $\theta_{01}$ and $\theta_{12}$ shown in Table~\ref{tab:theta}. Given that we are free to choose any $\Prob(\eps_0, \eps_1, \eps_2)$, the bounds in Eq.~(\ref{realbound}) and Eq.~(\ref{imbound}) follow immediately.

\begin{table}[]
\centering
\caption{Displays the various combinations of $\theta_{01}$ and $\theta_{12}$ in Eq.~(\ref{reim})}
\label{my-label}
\begin{tabular}{|c|c|c|c|}
\hline
\rowcolor[HTML]{FFFFFF} 
{\color[HTML]{000000} $\theta_{01}$} & {\color[HTML]{000000} $\theta_{12}$} & {\color[HTML]{000000} $\text{cos}(\theta_{12})+\text{cos}(\theta_{01})-\text{cos}(\theta_{12}+\theta_{01})$} & {\color[HTML]{000000} $\text{sin}(\theta_{12})+\text{sin}(\theta_{01})-\text{sin}(\theta_{12}+\theta_{01})$} \\ \hline
$0$                                  & $0$                                  & $1$                                                                                                          & $0$                                                                                                          \\ \hline
$0$                                  & $-\lambda \eps$                      & $1$                                                                                                          & $0$                                                                                                          \\ \hline
$0$                                  & $\lambda \eps$                       & $1$                                                                                                          & $0$                                                                                                          \\ \hline
$-\lambda \eps$                      & $+\lambda \eps$                      & $2\text{cos}(\lambda \eps)-1$                                                                                & $0$                                                                                                          \\ \hline
$\lambda \eps$                       & $-\lambda \eps$                      & $2\text{cos}(\lambda \eps)-1$                                                                                & $0$                                                                                                          \\ \hline
$-\lambda \eps$                      & $0$                                  & $1$                                                                                                          & $0$                                                                                                          \\ \hline
$\lambda \eps$                       & $0$                                  & $1$                                                                                                          & $0$                                                                                                          \\ \hline
\end{tabular}
\label{tab:theta}
\end{table}

\section{Maximal violations for the characteristic function}\label{sec:c}

Here we derive the quantum bounds for Eq.~(\ref{legmgf}). The moments of work obtained by successive projective measurements are given by Eq.~(\ref{momentsquantum}), and can be substituted into Eq.~(\ref{mgf}) by using 
\be
	G_{\lambda}(t_i,t_j)&=&\average{e^{i\lambda W(t_i,t_j)}}=\sum_{k=0}^{\infty}\frac{(i\lambda)^{k}}{k!}\average{W^{k}(t_i,t_j)}.
\ee 
where we have used the series expansion for the complex exponential function in terms of the work moments. This leads to an expression for the real part of the characteristic function;
\be\label{eq:realpart}
	\nonumber\mathcal{R}\text{e}(G_{\lambda}(t_i,t_j))&=&\sum_{k=0}^{\infty}\frac{(-1)^{k}}{(2k)!}\lambda^{2k}\average{W^{2k}(t_i,t_j)}, \\
	\nonumber&=&1+\frac{(1-\bs{a}_j\cdot\bs{a}_i)}{2}\sum_{k=1}^{\infty}\frac{(-1)^{k}}{(2k)!}(\lambda\eps)^{2k} \\
	&=&1+\frac{(1-\bs{a}_j\cdot\bs{a}_i)}{2}\big[\text{cos}(\lambda \eps)-1\big],	
\ee
Substituting Eq.~(\ref{eq:realpart}) into Eq.~(\ref{legmgf}) yields the following;
\be\label{eq:realpartbound1}
	\mathcal{R}\text{e}(L_{\lambda})&=&\mathcal{R}\text{e}\bigg(G_{\lambda}(t_1,t_2)+G_{\lambda}(t_0,t_1)-G_{\lambda}(t_0,t_2)\bigg) \\
	\nonumber&=&1+\frac{\big(\text{cos}(\lambda \eps)-1\big)}{2}\bigg[1+\bs{a}_2\cdot\bs{a}_0-\bs{a}_2\cdot\bs{a}_1-\bs{a}_1\cdot\bs{a}_0\bigg],
\ee
To bound Eq.~(\ref{eq:realpartbound1}) we denote $\bs{a}_{i+1}\cdot\bs{a}_i=\text{cos}(\phi_{i})$ and use the trigonometric relation $\bs{a}_{2}\cdot\bs{a}_0=\text{cos}(\phi_{0}+\phi_{1})$. This leads to the following inequality;
\be
	\mathcal{R}\text{e}(L_{\lambda}) \leq\frac{5}{4}-\frac{1}{4}\text{cos}(\lambda \eps),
\ee
where the upper bound is obtained by choosing $\theta_{10}=\theta_{21}=\pi/3$, as with the Leggett-Garg inequality for even moments of work in Eq.~(\ref{legeven2}). 

Turning to the imaginary part of Eq.~(\ref{mgf}) and using the expression Eq.~(\ref{momentsquantum}) for the work moments we find the following;
\be\label{eq:impart}
	\mathcal{I}\text{m}(G_{\lambda}(t_i,t_j))&=&\sum^{\infty}_{k=0}\frac{(-1)^{k}}{(2k+1)!}\lambda^{2k+1}\average{W^{2k+1}(t_i,t_j)}\\
	\nonumber&=&\frac{(\bs{r}\cdot\bs{a}_i)(\bs{a}_j\cdot\bs{a}_i-1)}{2}\text{sin}(\lambda \eps).
\ee
For the bounds on the imaginary term, we substitute Eq.~(\ref{eq:impart}) into Eq.~(\ref{legmgf}) and choose the same set of parameters that were applied to Eq.~(\ref{legodd2}), namely $\theta_{10}=\theta_{21}=\pi/2$, ending with Eq.~(\ref{qmgfim}).

\section{Maximum violations for full-counting statistics}\label{sec:d}

In this section we provide a derivation of Eq.~(\ref{eq:fcsreal}) and Eq.~(\ref{eq:fcsim}). We first begin with the characteristic function Eq.~(\ref{eq:fcs}) for a two-level system described by the Hamiltonian Eq.~(\ref{ham}); 
\be
	G^{\text{FCS}}_{\lambda}(t_i,t_j)=\Trace\big[\hatstate e^{-i\frac{\lambda}{4}\bs{a}_{i}\cdot \bs{\hatSig}}e^{i\frac{\lambda}{2}\bs{a}_{j}\cdot \bs{\hatSig}}e^{-i\frac{\lambda}{4}\bs{a}_{i}\cdot \bs{\hatSig}} \big]
\ee
We now expand the exponential operators using $e^{-i\lambda\bs{a}_{i}\cdot \bs{\hatSig}}=\text{cos}(\lambda)\hat{I}-i\text{sin}(\lambda)\bs{a}_{i}\cdot \bs{\hatSig}$;
\be
	e^{-i\frac{\lambda}{4}\bs{a}_{i}\cdot \bs{\hatSig}}e^{i\frac{\lambda}{2}\bs{a}_{j}\cdot \bs{\hatSig}}e^{-i\frac{\lambda}{4}\bs{a}_{i}\cdot \bs{\hatSig}}
	=\text{cos}^2(\lambda \eps/2)\hat{I}&-&\frac{i}{2}\text{sin}(\lambda \eps)(\bs{a}_{i}\cdot \bs{\hatSig})+i\text{sin}(\lambda \eps/2)\text{cos}^2(\lambda \eps/4)(\bs{a}_{j}\cdot \bs{\hatSig}) \\
	\nonumber &-&\frac{1}{2}\text{sin}^2(\lambda \eps/2)\big\lbrace (\bs{a}_{i}\cdot \bs{\hatSig}),(\bs{a}_{j}\cdot \bs{\hatSig})\big\rbrace -i\text{sin}^2(\lambda \eps/4)\text{sin}(\lambda \eps/2)(\bs{a}_{i}\cdot \bs{\hatSig})(\bs{a}_{j}\cdot \bs{\hatSig})(\bs{a}_{i}\cdot \bs{\hatSig})
\ee
Now we substitute this expansion back into $G^{\text{FCS}}_{\lambda}(t_i,t_j)$ above, along with $\hatstate=\frac{1}{2}(\hat{I}+\bs{r}\cdot\bs{\hatSig})$, and apply the identity $\Trace[(\bs{a}\cdot \bs{\hatSig})(\bs{b}\cdot \bs{\hatSig})]=2\bs{a}\cdot\bs{b}$;
\be\label{eq:g1}
	\nonumber G^{\text{FCS}}_{\lambda}(t_i,t_j)=\text{cos}^2(\lambda \eps/2)&-& \frac{i}{2}\text{sin}(\lambda \eps)(\bs{r}\cdot \bs{a}_{i})+i\text{sin}(\lambda \eps/2)\text{cos}^2(\lambda \eps/4)(\bs{r}\cdot \bs{a}_{j})+\text{sin}^2(\lambda \eps/2)(\bs{a}_{i}\cdot \bs{a}_{j})\\
	&+& \frac{1}{4}\text{sin}^2(\lambda \eps/2)\Trace\bigg[(\bs{r}\cdot \bs{\hatSig})\big\lbrace (\bs{a}_{i}\cdot \bs{\hatSig})(\bs{a}_{j}\cdot \bs{\hatSig})\big\rbrace\bigg] \\
	\nonumber&-&\frac{i}{2}\text{sin}^2(\lambda \eps/4)\text{sin}(\lambda \eps/2)\bigg(\Trace\bigg[(\bs{a}_{i}\cdot\bs{\hatSig}) (\bs{a}_{j}\cdot \bs{\hatSig})(\bs{a}_{i}\cdot \bs{\hatSig})\bigg]+\Trace\bigg[(\bs{r}\cdot\bs{\hatSig})(\bs{a}_{i}\cdot\bs{\hatSig}) (\bs{a}_{j}\cdot \bs{\hatSig})(\bs{a}_{i}\cdot \bs{\hatSig})\bigg]\bigg)
\ee
To simplify Eq.~(\ref{eq:g1}) we make use of the identity $(\bs{a}_{i}\cdot \bs{\hatSig})(\bs{a}_{j}\cdot \bs{\hatSig})=(\bs{a}_{i}\cdot \bs{a}_j)\hat{I}+i(\bs{a}_{i}\times \bs{a}_{j})\cdot \bs{\hatSig}$. This leads to the following set of relations;
\be
	&\Trace&\bigg[(\bs{r}\cdot \bs{\hatSig})\big\lbrace (\bs{a}_{i}\cdot \bs{\hatSig})(\bs{a}_{j}\cdot \bs{\hatSig})\big\rbrace\bigg]=2i\big[(\bs{a}_{i}\times \bs{a}_{j})\cdot \bs{r}+(\bs{a}_{j}\times \bs{a}_{i})\cdot \bs{r}\big]=0, \\
	&\Trace&\bigg[(\bs{a}_{i}\cdot\bs{\hatSig}) (\bs{a}_{j}\cdot \bs{\hatSig})(\bs{a}_{i}\cdot \bs{\hatSig})\bigg]=2i(\bs{a}_{j}\times \bs{a}_{i})\cdot \bs{a}_i=0, \\
	&\Trace&\bigg[(\bs{r}\cdot\bs{\hatSig})(\bs{a}_{i}\cdot\bs{\hatSig}) (\bs{a}_{j}\cdot \bs{\hatSig})(\bs{a}_{i}\cdot \bs{\hatSig})\bigg]=2(\bs{r}\cdot \bs{a}_{i})(\bs{a}_{j}\cdot \bs{a}_{i})-2(\bs{r}\times \bs{a}_{i})\cdot(\bs{a}_{j}\times \bs{a}_{i}).
\ee
Substituting these into Eq.~(\ref{eq:g1}), collecting the real and imaginary terms and simplifying leads to Eq.~(\ref{eq:fcsreal});
\be
	\mathcal{R}\text{e}\bigg(G^{\text{FCS}}_{\lambda}(t_i,t_j)\bigg)=1+\frac{(1-\bs{a}_j\cdot\bs{a}_i)}{2}\big[\text{cos}(\lambda \eps)-1\big] 
\ee
and similarly
\be\label{eq:fcsim1}
	\nonumber\mathcal{I}\text{m}\bigg(G^{\text{FCS}}_{\lambda}(t_i,t_j)\bigg)=\text{sin}(\lambda\eps/2)\text{cos}^2(\lambda\eps/4)(\bs{r}\cdot \bs{a}_{j})&-&\frac{1}{2}\text{sin}(\lambda\eps)(\bs{r}\cdot \bs{a}_{i}) \\
	&-&\text{sin}(\lambda\eps/2)\text{sin}^2(\lambda\eps/4)\bigg[(\bs{r}\cdot \bs{a}_{i})(\bs{a}_{j}\cdot \bs{a}_{i})-(\bs{r}\times \bs{a}_{i})\cdot(\bs{a}_{j}\times \bs{a}_{i})\bigg]
\ee
Noting that one can rewrite $(\bs{r}\times \bs{a}_{1})\cdot(\bs{a}_{2}\times \bs{a}_{1})=\text{tanh}(\beta \eps/2)\big[\text{cos}(\theta_{01}+\theta_{12})-\text{cos}(\theta_{01})\text{cos}(\theta_{12})\big]$, substituting Eq.~(\ref{eq:fcsim1}) into the Leggett-Garg equation Eq.~(\ref{legmgf}) leads to the following simple expression;
\be
	\mathcal{I}\text{m}(L^{FCS}_{\lambda})= 2\text{sin}^{2}\big(\lambda \eps/4\big)\text{sin}\big(\lambda \eps/2\big)\text{tanh}\big(\beta \eps/2\big)\bigg[\text{cos}(\theta_{01}+\theta_{12})-\text{cos}(\theta_{01})\text{cos}(\theta_{12})\bigg]	
\ee
with the upper bound Eq.~(\ref{eq:fcsim}) given by setting $\theta_{10}=\theta_{21}=\pi/2$.

\section{Maximum violations for the Margenau-Hill distribution}\label{sec:e}

In this section we provide details of Leggett-Garg violations found from the characteristic function Eq.~(\ref{eq:mh}). As noted in the main text, a work distribution based on the Margenau-Hill distribution was proposed in \cite{Allahverdyan2014c} to extend the definition of fluctuating to states with initial energy coherences. Firstly, note that the Margenau-Hill distribution can be used to define a joint probability for the successive energy outcomes $\eps_i$, $\eps_j$ at times $t_i$ and $t_j$ respectively \cite{Johansen2007};
\be
	\Prob\bigg(\eps_i=\pm\frac{\eps}{2},\eps_j=\pm\frac{\eps}{2}\bigg)=\Trace\big[\hatstate\hatP^{\pm}_{a_i}\star\hatP^{\pm}_{a_j}\big]. 
\ee
The corresponding characteristic function of work is then as follows;
\be\label{eq:gmh}
	\nonumber G^{\text{MH}}_{\lambda}(t_i,t_j)&=& \average{e^{i\lambda W(t_i,t_j)}} \\
	\nonumber &=& \sum_{\eps_i,\eps_j}\Prob(\eps_i,\eps_j)e^{i\lambda(\eps_j-\eps_i)}\\
	&=&\Trace\big[\hatstate e^{i\lambda\bs{a}_{j}\cdot \bs{\hatSig}}\star e^{-i\lambda\bs{a}_{i}\cdot \bs{\hatSig}}\big].
\ee 
Taking the expansion $e^{-i\lambda\bs{a}_{i}\cdot \bs{\hatSig}}=\text{cos}(\lambda)\hat{I}-i\text{sin}(\lambda)\bs{a}_{i}\cdot \bs{\hatSig}$, we obtain the following operator expression;
\be
	e^{i\lambda\bs{a}_{j}\cdot \bs{\hatSig}}\star e^{-i\lambda\bs{a}_{i}\cdot \bs{\hatSig}}=\text{cos}^2(\lambda \eps/2)\hat{I}+i\text{sin}(\lambda \eps)(\bs{a}_{j}-\bs{a}_{i})\cdot \bs{\hatSig}+\text{sin}^2(\lambda \eps/2)(\bs{a}_{j}\cdot \bs{\hatSig})
\ee
Substituting this into Eq.~(\ref{eq:gmh}) and applying the same identities used throughout Appendix~\ref{sec:d} gives an expression for the characteristic function;
\be\label{eq:gmh2}
	G^{\text{MH}}_{\lambda}(t_i,t_j)=\text{cos}^2(\lambda \eps/2)+\text{sin}^2(\lambda \eps/2)(\bs{a}_{j}\cdot \bs{a}_{i})+i\text{sin}(\lambda \eps)(\bs{a}_{j}-\bs{a}_{i})\cdot\bs{r}
\ee
We can immediately see that the real part of $G^{\text{MH}}_{\lambda}(t_i,t_j)$ is again equivalent to Eq.~(\ref{eq:fcsreal}). As for the imaginary term, a straightforward substitution of Eq.~(\ref{eq:gmh2}) into Eq.~(\ref{legmgf}) reveals that 
\be
	\mathcal{I}\text{m}(L^{MH}_{\lambda})=0,
\ee
for all choices of $\theta_{ij}$.

\end{document}